\documentclass[aps,column,superscriptaddress,footinbib]{revtex4}  
\usepackage{epstopdf}
\usepackage{graphicx}  
\usepackage{bm}        
\usepackage{braket}
\usepackage{exscale}                  
\usepackage[intlimits]{amsmath}       
\usepackage{amsfonts}
\usepackage{amssymb,amscd}
\usepackage{color}
\usepackage{hyperref}
\usepackage{subfigure}
\usepackage[normalem]{ulem}
\usepackage{slashed}
\usepackage{leftidx}
\usepackage{xcolor,soul}
\usepackage{txfonts}
\usepackage{charter}
\usepackage{dsfont}

\begin{document}

\title{Quantum partial coherence measures constructed from Fisher information}

\author{Dong-Ping Xuan\footnote{E-mail: 2230501014@cnu.edu.cn}}
\affiliation{School of Mathematical Sciences, Capital Normal University, Beijing 100048, China}
\author{Zhong-Xi Shen\footnote{E-mail: 18738951378@163.com}}
\affiliation{School of Mathematical Sciences, Capital Normal University, Beijing 100048, China}
\author{Wen Zhou\footnote{E-mail:2230501027@cnu.edu.cn}}
\affiliation{School of Mathematical Sciences, Capital Normal University, Beijing 100048, China}
\author{Hua Nan\footnote{E-mail: nanhua@ybu.edu.cn}}
\affiliation{Department of Mathematics, College of Sciences, Yanbian University, Yanji 133002, China}
\author{Shao-Ming Fei\footnote{E-mail: feishm@cnu.edu.cn}}
\affiliation{School of Mathematical Sciences, Capital Normal University, Beijing 100048, China}
\affiliation{Max-Planck-Institute for Mathematics in the Sciences, 04103 Leipzig, Germany}
\author{Zhi-Xi Wang\footnote{E-mail: wangzhx@cnu.edu.cn}\footnote{Corresponding author}}
\affiliation{School of Mathematical Sciences, Capital Normal University, Beijing 100048, China}

\begin{abstract}
\textbf{Quantum mechanics gives a new breakthrough to the field of parameter estimation. In the realm of quantum metrology, the precision of parameter estimation is limited by the quantum Fisher information. We introduce the measures of partial coherence based on (quantum) Fisher  information by taking into account the post-selective non-unitary parametrization process. These partial coherence measures present a clear operational interpretation by directly linking the coherence to the parameter estimation accuracy. Furthermore, we explore the distinctions between our partial coherence measure and the quantum Fisher information within the context of unitary parametrization. We provide an analytical  expression for the partial coherence measure of two-qubit states. We  elucidate  the operational significance of the partial coherence measures by establishing the connections between the partial coherence measures and quantum state discrimination.}
\end{abstract}
\maketitle

\vspace{.4cm}

\section{Introduction}\label{sect1}
Quantum coherence, which emerges from the superposition of quantum states, is considered a crucial physical resource ~\cite{TYM,JA,BCP} and is vital for the quantum information processing \cite{EJSM}, quantum metrology \cite{VSL}, quantum optics \cite{RJG}, nanoscale thermodynamics \cite{VG} and even quantum computation \cite{liang2022quantum,naseri2022entanglement}.

In the last few years, the groundbreaking work by Baumgratz et al.~\cite{BCP} has catalyzed a surge of interest in the study of coherence from an axiomatic and quantitative perspective. Researchers have particularly intensified their focus on the field within the realm of fully decoherent processes such as the von Neumann measurements. In \cite{LUO2017} Luo and Sun delve into the concept of partial coherence, which emerges in systems subjected to partial decoherence operations, exemplified by L{\"u}ders measurements. A genuine measure of partial coherence is introduced by taking into account that the the coherence of L{\"u}ders measurements is more general and versatile than that of the von Neumann measurements.

In order to address the fact that the K-coherence \cite{Girolami}, a measure for quantifying coherence based on the Wigner-Yanase skew information, does not satisfy the monotonicity under incoherent operation in the framework of coherence measure, Luo and Sun \cite{LUO2017} suggested a modified K-coherence by replacing the skew information based on observables with a version grounded in the spectral decomposition of these observables. This adjustment is motivated by the fact that the spectral decomposition of observables typically results in L{\"u}ders measurements, which capture only partial information of coherence. Consequently, they introduced the notion of partial coherence and the corresponding measure given by sum of the skew information across all measurement operators.

The concept of partial coherence has been instrumental in exploring the interplay among diverse quantum resources. Within multipartite quantum systems, there is an intricate link between coherence and quantum correlations \cite{XiZ,StreltsovA15}. It has been demonstrated that the quantum coherence of a bipartite state serves as a crucial resource in transforming a separable state into an entangled one \cite{Streltsov} or into a state with nonzero quantum discord \cite{JiajunMa}. In \cite{WangYTSCI}, by using entropy-based measures the authors investigated the relationships between quantum correlated coherence, the coherence between subsystems, and the quantum discord and the quantum entanglement.
In \cite{XiongPRA}, the correlated coherence in the framework of partial coherence theory is studied. It is shown that the partial coherence of a bipartite state is actually a measure of quantum correlation. In \cite{SLAJ} it is shown that that the presence of nonzero partial coherence is indispensable (necessary and sufficient) to generate quantum correlations via partial incoherent operations in a bipartite system. In \cite{SunhoKim2023} it is demonstrated  that the measures of partial coherence can be constructed via arbitrary symmetric concave functions, and have an explicit correspondence with the measures of entanglement.

Experimentally, instead of the state tomography, the authors in Ref.\cite{WangYT} presented a way to measure the coherence directly based on the interference fringes. In Ref.\cite{NajmehEtehadiAbari} the correlated quantum coherence of multiple quantum bits is studied. In Ref.\cite{WJZHAO} a quantum coherence protection scheme is proposed for a two-qubit system by taking into account the evolution of multi-interacting qubits in a common reservoir.

Physically, a bipartite state $\rho_{AB}$ associated with subsystems $A$ and $B$ is said to be $A$ coherent if its reduced state $\rho_A$ is coherent with respect to a given basis of the subsystem $A$. It implies that the whole state $\rho_{AB}$ may be coherent, i.e., not diagonal in the given bases, but the state $\rho_A$ is incoherent. Namely, the partial coherence is zero, while the global coherence is greater than zero.\cite{Parry2018}

Fisher information (FI) serves as a statistical measure that quantifies the amount of information a random sample holds about an underlying parameter. This concept is pivotal in statistical theory, guiding the design of experiments and the refinement of efficient estimators \cite{RAF}. The quantum Fisher information (QFI) extends the classical notion and provide a quantum mechanical adaptation that is essential in quantum metrology and quantum information theory \cite{CMCaves1996,Braunstein1996}. The classical FI evaluates the variance of an estimator, whereas quantum Fisher information (QFI) quantifies the variance in the evolution of quantum states. This distinction implies that QFI incorporates the nuances of quantum entanglement and coherence, the phenomena absent in the classical context. The invariance of QFI under unitary transformations sets it apart from its classical counterpart. This characteristic stems from the intrinsic probabilistic nature of quantum mechanics, which does not permit deterministic state transformations. Moreover, due to the complex Hilbert space structure of quantum states, QFI can assume negative values, unlike the classical version that is strictly non-negative. These disparities underscore the unique challenges and opportunities presented by quantum information processing.

QFI serves as a fundamental concept in quantum metrology, representing the limit of precision in parameter estimation from a quantum state, referred to as the Cram\'{e}r-Rao  bound \cite{SC}. Beyond its significance in metrology, QFI has found applications in the description of criticality and quantum phase transitions \cite{LP}, establishing speed limits for quantum computation \cite{SL}, as well as identifying quantum entanglement in composite systems \cite{NS}. QFI and Wigner-Yanase skew information are recognized as extensions of classical Fisher information, with QFI being the primary focus in modern studies. Hence, a natural problem arises in quantifying the coherence directly by using QFI \cite{HKS}.

The coherence of the probing state is often a crucial component in various quantum metrology processes \cite{VG,VSL2011,VSL2006}. For example, when performing the usual phase estimation of a parameter $\eta$ using the unitary parametrization $\mathcal{U}_\eta(\cdot)=e^{-i\eta H}(\cdot)e^{i\eta H}$, it is essential for the state to exhibit coherence relative to the eigenvectors of the Hermitian operator $H$. Furthermore, the state that exhibits the maximum coherence in the sequential protocol has the potential to realize the optimal estimation of an unknown parameter\cite{VSL2006}. One crucial problem in quantum metrology is the QFI, which sets a limit on the estimation precision.
\cite{AVDA,RAF,MH,SC}. A straightforward calculation demonstrates that the QFI, under unitary parametrization $\mathcal{U}_\eta(\cdot)$ in the qubit scenario \cite{WZJ}, varies monotonically with certain coherence measures such as $l_1$ norm coherence.
Numerous studies have explored the connections between the quantum coherence and the Fisher information (FI) or QFI \cite{SLAJ,SY,BPCD,BV,KVB,LQSM,KSH,XLF,TMA,HKSH}. In certain contexts, coherence can be elucidated by QFI (or FI) \cite{BPCD,BV,TMA}. Importantly, QFI under unitary parametrization is closely linked to inexpressible coherence \cite{ BV,IMRW}, which is a specific aspect of the resource theory of asymmetry \cite{STR,GR,IR}.
In addition, coherence measures are defined as free operations based on QFI with respect to the dephasing parameter in the context of strictly incoherent operations \cite{BPCD}. However, the quantification of quantum partial coherence in general scenarios has not yet been done directly using the estimation accuracy and FI (or QFI). An issue is that in the general resource theory of coherence, QFI with the unitary parametrization $\mathcal{U}_\eta(\cdot)$ is not a coherence measure \cite{BCP}. In one case, under incoherent operations, 3-dimensional maximally coherent states could be transformed into 2-dimensional maximally coherent states \cite{KVB,AD,SZX}. Therefore, in the subsequent work \cite{YY} the authors identified a suitable parameterization process to establish coherence measures for single quantum states and examined how coherence functions in quantum metrology.

The main topic of this study is the quantification of partial quantum coherence, which is an extension of normal coherence \cite{SLAJ,LUO}. Subject to a kind of non-unitary parametrization, we define several equivalent measures of partial coherence in terms of the general resource theory of partial coherence provided by the FI (and QFI).
Through the optimal estimation accuracy with non-unitary parametrization, our partial coherence measure naturally inherits the operational meaning of FI, as the optimal estimation accuracy is constrained by FI, which is asymptotically obtained with maximum likelihood estimators \cite{AVDA,RAF,MH}. We further demonstrate that in the case of two-qubit, we present the analytic expression of our partial coherence measure and show that it can be equivalently understood under unitary parametrizations. In addition to establishing a clear relationship between coherence and parameter estimating accuracy (or FI), our partial coherence measure also clarifies the functions of the non-unitary parametrization process. We also provide an operational interpretation for the partial coherence measures based on QFI from quantum state discrimination (QSD).

The paper is organized as follows. We review the fundamental concepts of resource theory of partial coherence in Sec.~\ref{sect2}. We also give a brief summary of the parametrization process and define FI and QFI in terms of parametrization. We next present the main results for partial coherence measures constructed from FI in Sec.~\ref{sect3}. We give the analytical conclusions of the partial coherence measure for the two-qubit scenario in Sec.~\ref{sect4}, along with the partial coherence measures based on QFI. The equivalency with the unitary parametrization is also examined. We present detailed examples to demonstrate our results. We establish the relationships between QFI, QFI partial coherence measure, and QSD in Sec.~\ref{sect5}. We conclude and discuss in Sec.~\ref{sect6}.

\vspace{.6cm}
\section{Partial coherence and Fisher information}\label{sect2}
\noindent
Consider bipartite systems in Hilbert space $H^{AB}=H^A\otimes H^B$, where the Hilbert spaces of the subsystems $A$ and $B$ have finite dimensions $n_A$ and $n_B$, respectively, and are denoted by the Hilbert spaces $H^A$ and $H^B$.

Let $\Pi_L=\left\{\Pi_i^A \otimes \mathbf{1}^B\right\}$ be the L{\"u}ders measurement extension of a fixed local von Neumann measurement $\Pi^A=$ $\left\{\Pi_i^A\right\}$ on party $A$. Then the notion of ``partial coherence" with respect to the L{\"u}ders measurement $\Pi_L$ is given as follows \cite{SY}:

(i) The set of partial incoherent states is defined by
$$
\mathcal{I}_P^a=\left\{\sigma: \Pi_L\left(\sigma\right)=\sigma\right\},
$$
where $\Pi_L\left(\sigma\right)=\sum_i\left(\Pi_i^A \otimes \mathbf{1}^B\right) \sigma\left(\Pi_i^A \otimes \mathbf{1}^B\right)$.

For a fixed basis $\{\ket{i}:\,i=1,2,3,...,n_A\}$ on party $A$,
the partial incoherent state, i.e., the free state in the resource theory of
partial coherence, can be written as
\begin{align}\label{eq22}
\sigma=\sum_ip_i\ket{i}\bra{i}\otimes\sigma_i,
\end{align}
where $p_i=\mathrm{tr}(\ket{i}\bra{i}\otimes I^B)\,\sigma\,(\ket{i}\bra{i} \otimes I^B)$ and $\sigma_i={\mathrm{tr_A}(\sigma)}/{p_i}$, with \(\text{tr}_A \) denoting the partial trace over system A.

(ii) A completely-positive and trace-preserving (CPTP) map $\Phi^A$ with Kraus operators $\{K_n\}$ is referred to as partial incoherent if $K_nI^A_PK^{\dagger}_n\in I^A_P$. The set of partial incoherent operations (PIO) is denoted as $\mathcal{O}^A_P$.

In the context of a bipartite state $\rho^{AB}\in H^{AB}$ undergoing a quantum channel $\Phi_\varepsilon$ depending on a parameter $\varepsilon$, we aim to estimate the unknown parameter through measurements on $\Phi_\varepsilon(\rho^{AB})$. Our focus here is on the free parametrization processes of $\Phi_\varepsilon=\{K^A_l(\varepsilon)\otimes I^B\}$,
\begin{align}
\Phi_l(\varepsilon)=K^A_l(\varepsilon)\otimes I^B=\sum_n{a_n^l(\varepsilon)}|f_l(n)\rangle\langle n|\otimes I^B,
\end{align}
where $K^A_l(\varepsilon)=\sum_na_n^l(\varepsilon)|f_l(n)\rangle\langle n|$ is the Kraus on subsystem $A$, $\sum_lK^A_l(\varepsilon)^\dagger K^A_l(\varepsilon)=I^A$, $\{|n\rangle\}$ is the fixed basis, $f_l(\cdot)$ maps from integers to integers, $I^A$ ($I^B$) is the identity in system $A$ ($B$).

We aim to ensure that the presence of a partial incoherent probe throughout the parametrization process does not have any impact on the parameter estimation, illustrating the importance of the partial coherence. In other words, the parameter $\varepsilon$ should not affect the measurement results $\Phi_\varepsilon(\sigma)$ and $\{\sigma_l, p_l \}$ obtained from a partial incoherent probe $\sigma\in I^A_P$, where $p_l=\operatorname{tr}[\Phi_l(\varepsilon)\sigma \Phi_l(\varepsilon)^\dagger]$ and $\sigma_l=\Phi_l(\varepsilon)\sigma \Phi_l(\varepsilon)^\dagger/p_l$.
As a result, $|a_n^l(\varepsilon)|$ is independent of parameter $\varepsilon$, and  $\Phi_l(\varepsilon)$  can be expressed as
\begin{align}
\Phi_l(\varepsilon)=K^A_l(\varepsilon)\otimes I^B=\sum_nb_n^le^{ig_n^l(\varepsilon)}|f_l(n)\rangle\langle n|\otimes I^B,\label{dy}
\end{align}
where $g_n^l$ is a real function, and $b_n^l$ is parameter-independent.
Actually, it shares a similarity with the conventional phase estimation $\mathcal{U}_\varepsilon(\cdot)=e^{-i \varepsilon H}(\cdot)e^{i\varepsilon H}$. The phase estimation process's measurement results from a partial incoherent probe are independent of $\varepsilon$. Furthermore, similar to (\ref{dy}), the unitary operator $\mathcal{U}_\varepsilon$ can be expressed in terms of the observable $e^{-iH\varepsilon}=\sum_ne^{-ih_n\varepsilon}|n\rangle\langle n|$, where $h_n$ is eigenvalue of $H$. From this point of view, one can see the parametrization process $\Phi_\varepsilon$ as a non-unitary version of the unitary phase estimation.

Denote $G$ the operations given in Eq.(\ref{dy}) with $\partial_\varepsilon g_n^l(\varepsilon)\in[0,1]$ (see Appendix G for the detailed derivation of the range) and $G_1$ the PIO satisfying $Rank\left [K^A_l(\varepsilon)^\dagger K^A_l(\varepsilon)\right]=1$.

The probability distribution can be obtained directly from the PIO $\Phi_\varepsilon$ on a quantum bipartite state $\rho^{AB}$ if post-selection is permitted. We can use the following equation to represent:
\begin{align}
P^{\scriptscriptstyle\Phi}(l|\varepsilon)=&\operatorname{tr}(\Phi_l(\varepsilon)\rho^{AB} \Phi_l(\varepsilon)^\dagger)\nonumber\\
=&\operatorname{tr}((K^A_l(\varepsilon)\otimes I^B)\rho^{AB} (K^A_l(\varepsilon)\otimes I^B)^\dagger).\label{post}
\end{align}
$\Phi_\varepsilon(\rho^{AB})$ will be the state following the PIO if post-selection is not permitted. The probability distribution family is given by
\begin{align}
P_{\scriptscriptstyle\mathcal{M}}^{\scriptscriptstyle\Phi}(l|\varepsilon)=\operatorname{tr}((M^A_l\otimes I^B)\Phi_\varepsilon(\rho^{AB})),\label{gen}
\end{align}
where the subscript $\mathcal{M}$ denotes the general measurement can be obtained by operating a measure $\mathcal{M}=\{M^A_l\otimes I^B\}$ on the state $\Phi_\varepsilon(\rho^{AB})$, where $\{M^A_l\}$ is the positive operator value measure (POVM) on the subsystem $A$.


Originated from the statistical inference, the total quantity of information that an observable random variable $X$ carries about an unknown parameter $\mu$ is measured by the classical Fisher information \cite{RAF,WY}. Assume that the probability density $\left\{p_i(\mu), \mu\in\right.$ $\mathbb{R}\}_{i=1}^N$is conditioned on a fixed value of the parameter $\mu=\mu^*$, and that the measurement results are $\left\{x_i\right\}$. The classical Fisher information is given by
\begin{align}
F_\mu=\sum_i p_i(\mu)\left[\frac{\partial \ln p_i(\mu)}{\partial \mu}\right]^2.\label{FI}
\end{align}
It describes the maximum-likelihood estimator's inverse variance of its asymptotic normalcy. In this case, it is assumed that the observable is discrete, otherwise, an integral should be used in place of the summation in Eq.(\ref{FI}).

From the above parameterization process, we define the Fisher information of distribution $P^{\scriptscriptstyle\Phi}(l|\varepsilon)$ by
\begin{align}
F(P^\Phi,\varepsilon_0)=\sum_lP^\Phi(l|\varepsilon_0)\left[\left.\frac{\partial \ln P^\Phi(l|\varepsilon)}{\partial\varepsilon}\right|_{\varepsilon_0}\right]^2,\label{Fisher}
\end{align}
then the quantum Fisher information  of $P_{\scriptscriptstyle\mathcal{M}}^{\scriptscriptstyle\Phi}(l|\varepsilon)$ defined by
\begin{align}
F_{\scriptscriptstyle Q}(\rho^{AB},\Phi,\varepsilon_0)=\mathop{\max}\limits_{\mathcal{M}}
F(P_{\scriptscriptstyle\mathcal{M}}^{\scriptscriptstyle\Phi},\varepsilon_0),\label{QFI}
\end{align}
for any given $\varepsilon_0$.

\vspace{.4cm}
\section{Partial coherence measures constructed from FI}\label{sect3}
\noindent
Based on the partial incoherent states and partial incoherent operations, the resource theory of partial coherence can be established similarly to the usual coherence\cite{LUO2017,XiongPRA}. A functional $C^A$ is a measure of partial coherence in relation to the L{\"u}ders measurement $\Pi_L$, for bipartite systems. The functional  $C^A$ satisfies the following conditions (C1-C4),
\begin{itemize}
\item[(C1)] \textit{Nonnegativity}: $C^A(\rho^{AB})\ge0$, with the equality holding if and only if $\rho^{AB}\in I^A_P$.

\item[(C2)]\textit{Monotonicity under partial incoherent operations}: $C^A(\Phi^A(\rho^{AB}))\le C^A(\rho^{AB})$ for all $\Phi^A\in\mathcal{O}^A_P$.

\item[(C3)]\textit{Monotonicity under selective partial incoherent operations on average}: $\sum_ip_iC^A(p^{-1}_iK_i\rho^{AB}K^{\dagger}_i)\le C^A(\rho^{AB})$ with  $p_i=\mathrm{tr}(K_i\rho^{AB}K^{\dagger}_i)$ the probabilities and $K_i$ the partial incoherent Kraus operators.

\item[(C4)]\textit{Convexity}:  $C^A(\sum_ip_i\rho^{AB}_i)\le \sum_ip_iC^A(\rho^{AB}_i)$ for any ensemble $\{p_i, \rho^{AB}_i\}$ with $p_i\ge0$ and $\sum_ip_i=1$.
\end{itemize}

From the Fisher information, we have the following partial coherence measures, as shown by the theorems below. (The specific proof process is in the appendix.)

\textbf{Theorem 1}.
For a bipartite state $\rho^{AB}$, the partial coherence  $C^{\scriptscriptstyle\varepsilon_0}_F(\rho^{AB})$  can be quantified through the maximum FI for a given parameter $\varepsilon_0$,
\begin{align}
C^{\scriptscriptstyle\varepsilon_0}_F(\rho^{AB})=\mathop{\max}\limits_{\Phi\in G}
F(P^{\scriptscriptstyle\Phi},\varepsilon_0),\label{d22}
\end{align}
where $P^{\scriptscriptstyle\Phi}$ is the distribution (\ref{post}) and $F(P^{\scriptscriptstyle\Phi},\varepsilon_0)$  is the FI given by (\ref{Fisher}).

In the context of incoherent non-unitary parametrization, we may observe from Theorem 1 that the partial coherence can be defined by the FI of the probability distribution (\ref{post}), indicating a link between coherence and the estimate accuracy. Actually, $G_1$, the subset of $G$, can be used in place of $G$ in the definition (\ref{d22}). This can be represented by Lemma 1 below, and the proof process is shown in the Appendix.

\textbf{Lemma 1}. There always exists a $\Phi'=\{\tilde{K}_{l}(\varepsilon)\otimes I^B\}\in G_1$ such that $F(P^{\scriptscriptstyle\Phi},\varepsilon_0)\leq F(P^{\scriptscriptstyle\Phi'},\varepsilon_0)$, for each $\Phi=\{K_l(\varepsilon)\otimes I^B\}\in G$, where $F(P^{\scriptscriptstyle\Phi},\varepsilon_0)$ and
$F(P^{\scriptscriptstyle\Phi'},\varepsilon_0)$ are FI of $P^{\scriptscriptstyle\Phi}(l|\varepsilon)$ and $P^{\scriptscriptstyle\Phi'}(l|\varepsilon)$, respectively, $\{K_l(\varepsilon)\}$ and $\{\tilde{K}_{l}(\varepsilon)\}$ are the Kraus operators on subsystem $A$.

Lemma 1 implies that optimizing over the subset $G_1$ will maximize the FI over the set G, hence reducing the range of the optimal PIO.

\vspace{.4cm}
\section{Partial coherence measures constructed from QFI}\label{sect4}
\noindent
The FI related to the probability distribution produced by post-selective PIO on a bipartite state is the main focus of Theorem 1. Before we consider the measure of partial coherence based on QFI with respect to the parametrization in $G$, we present the following Lemma 2. (The specific proof process is in the Appendix.)

\textbf{Lemma 2}. The FI induced directly by the optimal post-selective PIO parametrization process,
\begin{align}
\mathop{\max}\limits_{\Phi\in
G}F_{\scriptscriptstyle Q}(\rho^{AB},\Phi,\varepsilon_0)\leq\mathop{\max}\limits_{\Phi\in G}
F(P^{\scriptscriptstyle\Phi},\varepsilon_0),\label{QFI2}
\end{align}
sets an upper limit on the maximum QFI when the parametrization is done in $G$, where $P^{\scriptscriptstyle\Phi}$  is the distribution mentioned in (\ref{post}).

By using the above Lemma 2, we have the following theorem. (The specific proof process is in the Appendix.)

\textbf{Theorem 2}.
The following partial coherence measure, as obtained from QFI respecting the parametrization in $G$,
\begin{align}
C_{\scriptscriptstyle Q}^{\scriptscriptstyle\varepsilon_0}(\rho^{AB})=\mathop{\max}\limits_{\Phi\in G}F_{\scriptscriptstyle Q}(\rho^{AB},\Phi,\varepsilon_0)\label{CQFI}
\end{align}
is equivalent to the measure of partial coherence defined by FI,
$C_{\scriptscriptstyle Q}^{\scriptscriptstyle\varepsilon_0}(\rho^{AB})
=C^{\scriptscriptstyle\varepsilon_0}_F(\rho^{AB})$.

We have demonstrated the equivalence of the partial coherence measures based on FI and QFI under post-selective parametrization. This kind of partial coherence measure's primary benefit is that it can be directly linked to the parameter estimate process via the Cram\'{e}r-Rao bound \cite{RAF,MH,SC,RCR}. In the following, we demonstrate how our partial coherence measure can be related to some quantum metrology schemes. This also provides an operational meaning of our partial coherence measure.

The scheme is described as follows. Suppose a bipartite state $\rho\in H^{AB}$ undergoes a quantum channel $\Phi_\varepsilon$ depending on a parameter $\varepsilon$. This endows an unknown phase $\varepsilon$ to the state $\rho$ as $\rho_\varepsilon=\Phi_\varepsilon \rho {\Phi_\varepsilon}^{\dagger}$. We aim to estimate $\varepsilon$ in $\rho_\varepsilon$ by $N>>1$ runs of detection on $\rho_\varepsilon$. Based on the quantum parameter estimation\cite{RAF},
the mean square error   $\Delta{\varepsilon}$ in the estimation of $\varepsilon$ is bounded below by the quantum Cram\'er-Rao bound (CRB),
$$
\left(\Delta{\varepsilon}\right)^2 \geqslant \frac{1}{N F_{Q}},
$$
where $F_{Q }$ is the quantum Fisher information, and $N$ the number of repeated experiments.

The CRB is directly related to the inverse of the QFI, hence linking the information content of the quantum state to the precision of parameter estimation. In fact, since the variance $\Delta{\varepsilon}$ usually deviates from the optimal one $\Delta{\varepsilon'}$, one has $\Delta{\varepsilon} \geqslant \Delta{\varepsilon'}$. That is, the larger the $F_{Q }$, the smaller the variance $\Delta{\varepsilon}$ and the closer to $\Delta{\varepsilon'}$. The definition of a partial coherence measure based on QFI reveals that $(\Delta{\varepsilon})^2$ is bounded by our coherence $C_{\scriptscriptstyle Q}^{\scriptscriptstyle\varepsilon_0}=\mathop{\max}\limits_{\Phi\in G}F_{\scriptscriptstyle Q}$.

The QFI quantifies the amount of information that a quantum state carries about a parameter of interest. The CRB establishes a lower bound on the variance of any unbiased estimator of a parameter. The above result implies that the partial coherence measure has operational significance in the context of quantum metrology. Its connection to the CRB also provides some physical insight into the role of coherence played in the parameter estimation. It suggests that coherence can be harnessed to improve the precision of measurements. This insight can guide the design of quantum sensors and metrological protocols. By knowing how the partial coherence relates to the parameter estimation precision, one can optimize the protocols to maximize the QFI, by engineering the system's coherence or choosing optimal measurement bases to enhance the partial coherence and thus the precision of the parameter estimation.

While the partial coherence measures carry a clear operational interpretation in quantum metrology, finding their analytically computable expressions is difficult. In the following theorem, we demonstrate that for a quantum state in two-qubit system, we can obtain an analytic result and realize the partial coherence measure using QFI with unitary parametrization. (The specific proof process is in the Appendix.)

\textbf{Theorem 3}.
The partial coherence for any $2\otimes2$ state $\rho^{AB}$ is provided by
\begin{align}
C^{\scriptscriptstyle \varepsilon_0}_F(\rho^{AB})=F_{\scriptscriptstyle Q}(\rho^{AB},U_\varepsilon,\varepsilon_0),
\end{align}
here $F_{\scriptscriptstyle Q}$ denotes the unitary parametrization $U_\varepsilon=e^{i\varepsilon}|1\rangle\langle 1|\otimes I^{B}+|2\rangle\langle 2|\otimes I^{B}$  constrained by the QFI  of $\rho^{AB}$.

Theorem 3 shows that for two-qubit case, our partial coherence measure is directly related to the one based on unitary parametrization. Actually, for the general high-dimensional instance, the FI with unitary parametrization is not equivalent to $C^{\scriptscriptstyle\varepsilon_0}_F$. We illustrate their differences by specific examples.

\textbf{Example 1}.  Consider a bipartite state $\rho_1=|\phi_A\rangle\otimes|\varphi_B\rangle$ with $|\phi\rangle_A=(\frac{1}{\sqrt{2}},\frac{1}{2},\frac{1}{2})^T$ and $|\phi\rangle_B=(\frac{1}{\sqrt{2}},0,\frac{1}{\sqrt{2}})^T$.

The FI is as follows
\begin{align*}
F(P^\Phi,0)=\sum_l\frac{[\partial_\varepsilon P(l|\varepsilon)|_0]^2}{P(l|0)}=1.07.
\end{align*}
So we note from definition (\ref{d22}), $C^0_F(\rho_1)\geq F(P^\Phi,0).$

In order to compare our measure with QFI, taking into account the optimal unitary parametrization in $G$, we compute
\begin{align*}
\mathop{\max}\limits_{U_\varepsilon\in G}F_{\scriptscriptstyle Q}(|\varphi\rangle,U_\varepsilon,0)=&\mathop{\max}\limits_{H\in S}4\langle\varphi|H^2_{\varepsilon_0}|\varphi\rangle-4\langle\varphi|H_{\varepsilon_0}|\varphi\rangle^2=1.00.
\end{align*}

Thus $C^0_F(\rho_1)$ exceeds the maximum value of $F(|\varphi\rangle,U_\varepsilon,0)$, indicating that $C^{\scriptscriptstyle\varepsilon_0}_F$  is distinct from the FI with unitary parametrization, see Appendix F for detailed derivations.

\textbf{Example 2}.  Consider a bipartite state $\rho_2=|\phi_A\rangle\otimes|\phi_B\rangle$, where $|\phi\rangle_A=(\frac{1}{3},\frac{2}{3},\frac{2}{3})^T$ and $|\phi\rangle_B=(\frac{1}{\sqrt{2}},\frac{1}{2},\frac{1}{2})^T$. Similar to the calculations in Example 1, we obtain the corresponding FI,
\begin{align*}
F(P^\Phi,0)=\sum_l\frac{[\partial_\varepsilon P(l|\varepsilon)|_0]^2}{P(l|0)}=0.71.
\end{align*}
From the definition (\ref{d22}), since
\begin{align*}
\mathop{\max}\limits_{U_\varepsilon\in G}F_{\scriptscriptstyle Q}(|\varphi\rangle,U_\varepsilon,0)=&\mathop{\max}\limits_{H\in S}4\langle\varphi|H^2_{\varepsilon_0}|\varphi\rangle-4\langle\varphi|H_{\varepsilon_0}|\varphi\rangle^2=0.40,
\end{align*}
we have $C^0_F(\rho_2)>\mathop{\max}\limits_{U_\varepsilon\in G}F(|\varphi\rangle,U_\varepsilon,0)$. Thus, $C^0_F(\rho_2)\geq F(P^\Phi,0)$,

\textbf{Example 3}. Consider a bipartite state $\rho_3=|\phi_A\rangle\otimes|\phi_B\rangle$, where $|\phi\rangle_A=(\frac{1}{\sqrt{3}},\frac{1}{\sqrt{3}},\frac{1}{\sqrt{3}})^T$ and $|\phi\rangle_B=(\frac{1}{\sqrt{2}},0,\frac{1}{\sqrt{2}})^T$. The FI reads
\begin{align*}
F(P^\Phi,0)=\sum_l\frac{[\partial_\varepsilon P(l|\varepsilon)|_0]^2}{P(l|0)}=0.94.
\end{align*}
Since
\begin{align*}
\mathop{\max}\limits_{U_\varepsilon\in G}F_{\scriptscriptstyle Q}(|\varphi\rangle,U_\varepsilon,0)=&\mathop{\max}\limits_{H\in S}4\langle\varphi|H^2_{\varepsilon_0}|\varphi\rangle-4\langle\varphi|H_{\varepsilon_0}|\varphi\rangle^2=0.89,
\end{align*}
we have $C^0_F(\rho_3)\geq F(P^\Phi,0)$.

It is obvious from the previously mentioned examples that, in higher-dimensional instances, $C^{\scriptscriptstyle\varepsilon_0}_F$  differs from the FI with unitary parametrization. Fig. 1 makes the differences between the two easier to understand.

\begin{figure}
  \centering
  \includegraphics[width=18cm]{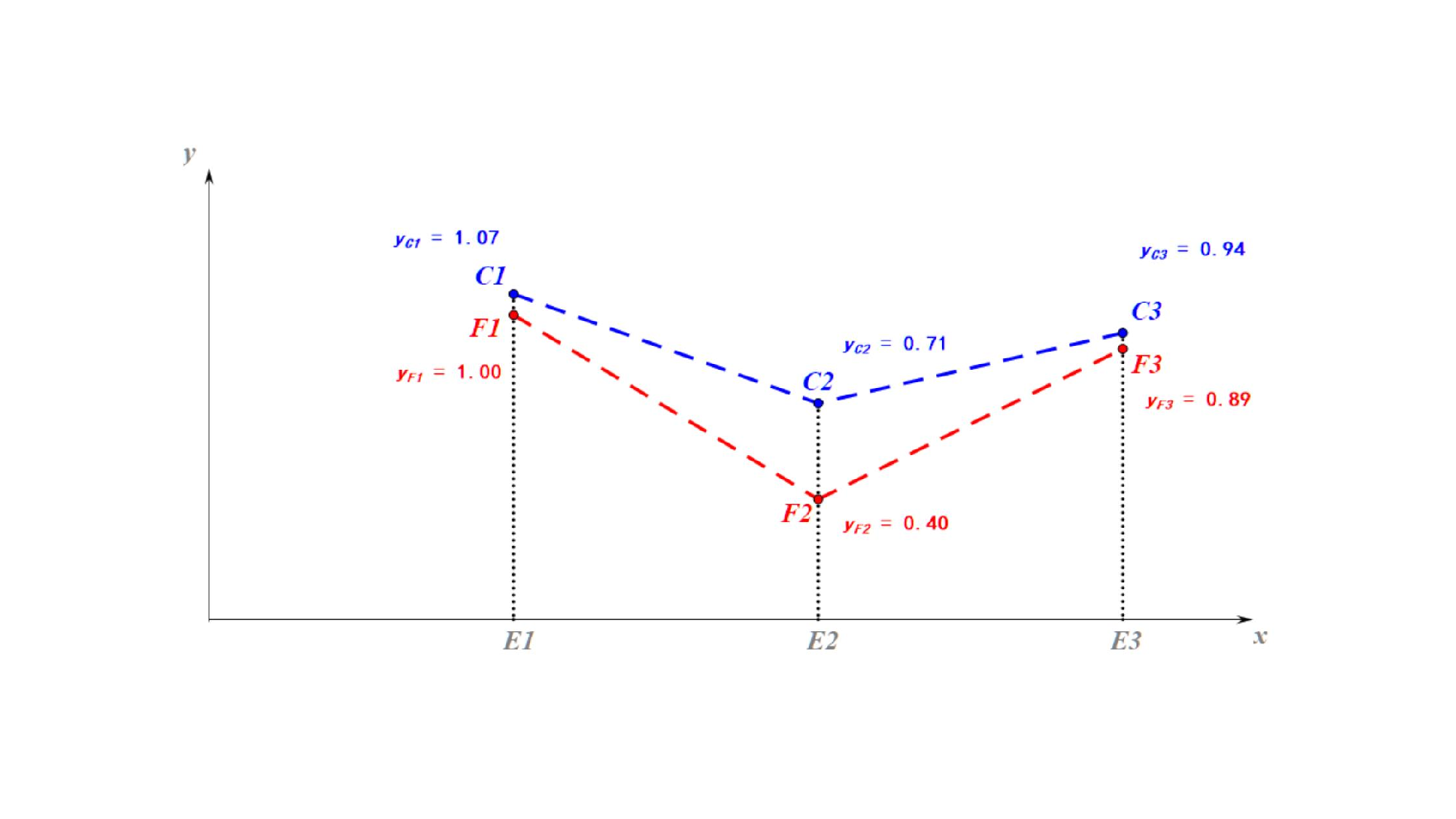}\\
  \caption{\small (Color online). The value of $C^0_F(\rho)$ is represented by the blue dashed line, while the value of $\mathop{\max}\limits_{U_\varepsilon\in G}F(|\varphi\rangle,U_\varepsilon,0)$ is represented by the red dashed line, the horizontal coordinates $E1,E2,E3$ stand for the quantum states $\rho_1,\rho_2$ and $\rho_3$ of the three examples, respectively.}
  \label{Example}
\end{figure}

\vspace{.4cm}
\section{QSD  based on parameterization process}\label{sect5}
\noindent

In QSD tasks, the receiver's goal is to determine the received state with maximum probability after the sender chooses a random state from the ensemble $\{\rho_i, \eta_i \}$.  In order to do this, the receiver performs a POVM $\{M_i : M_i \ge 0, \sum_i M_i =I\}$ on each $\rho_i$, and after reading $j$ from the measurement out, it declares the state to be $\rho_j$.

As the probability to get the result $j$ is $p_{j|i}=\mathrm{tr}(M_j\rho_i)$ when the system is in the state $\rho_i$, the maximal success probability to identify $\{\rho_i,\eta_i\}$ is
\begin{align}
P^{\mathrm{opt}}_\mathrm{S}(\{\rho_i,\eta_i\})=\mathop{\mathrm{max}}_{\{M_i\}}
\sum_i\eta_i\mathrm{tr}(M_i\rho_i),\nonumber
\end{align}
where the maximization goes over all POVMs $\{M_i\}$, while minimal error probability is
\begin{align}
P^{\mathrm{opt}}_\mathrm{E}(\{\rho_i,\eta_i\})
=1-\mathop{\mathrm{max}}_{\{M_i\}}\sum_i\eta_i\mathrm{tr}(M_i\rho_i).\nonumber
\end{align}


In QSD tasks, POVMs have a well-established ability to outperform von Neumann measurements \cite{AW1991,AF1990}. So far, however, the  optimal measurement for discriminating between a set of linearly independent pure and mixed states \cite{YE2003} has been the von Neumann measurement. The equivalency between the coherence theory and the pure state QSD task has been demonstrated in \cite{XIONG2018}. It is challenging to establish this kind of equivalency for general mixed states. However, the authors of \cite{XiongPRA} showed that the QSD task of mixed states only corresponds to the partial coherence by establishing a QSD state for each QSD task. We next use this result to establish the connection between QSD and QFI partial coherence.

For a state discrimination task $\left\{\rho_i, \eta_i\right\}_{i=1}^n$, where  $\rho_i$ is an $m \times m$ density matrix, let us consider an $m n \times m n$ matrix $\rho$ whose $(i, j)$-th entry is a block given by $\rho_{i j}=\sqrt{\eta_i \rho_i} \sqrt{\eta_j \rho_j}$, $1 \leq i, j \leq n$, that is,
\begin{equation}
\begin{aligned}
\rho=\left(\begin{array}{cccc}\eta_1 \rho_1 & \sqrt{\eta_1 \rho_1} \sqrt{\eta_2 \rho_2} & \cdots &\sqrt{\eta_1 \rho_1} \sqrt{\eta_n \rho_n} \\\sqrt{\eta_2 \rho_2} \sqrt{\eta_1 \rho_1} & \eta_2 \rho_2 & \cdots & \sqrt{\eta_2 \rho_2} \sqrt{\eta_n \rho_n} \\\cdot & \cdot & \cdots & \cdot \\\cdot & \cdot & \cdots & \cdot \\\sqrt{\eta_n \rho_n} \sqrt{\eta_1 \rho_1} & \sqrt{\eta_n \rho_n} \sqrt{\eta_2 \rho_2} & \cdots & \eta_n \rho_n\end{array}\right)\label{cccc}.
\end{aligned}
\end{equation}
As $\rho=A^{\dagger} A$, where $A=\left(\sqrt{\eta_1 \rho_1}, \sqrt{\eta_2 \rho_2}, \ldots, \sqrt{\eta_n \rho_n}\right)$, so it is positive semidefinite. Moreover, $\operatorname{tr} \rho=\sum_i \operatorname{tr}\left(\eta_i \rho_i\right)=\sum_i \eta_i=1$. Hence, $\rho$ is a density matrix. We call the state (\ref{cccc}) the QSD-state of $\left\{\rho_i,\eta_i\right\}_{i=1}^n$. Note that the corresponding QSD ensemble of $\rho$ is $\left\{\omega_i, p_i\right\}_{i=1}^n$, where $p_i=\operatorname{tr} \sqrt{\rho}|i\rangle\langle i|\otimes I_m \sqrt{\rho}$ and $\omega_i=p_i^{-1} \sqrt{\rho}|i\rangle\langle i| \otimes I_m \sqrt{\rho}$.

It is possible to find a $m \times m n$ unitary matrix $U$  such that $A=U \sqrt{\rho}$ by applying the polar decomposition theorem \cite{XiongPRA}. Consequently, for every $1 \leq i \leq n$,
\begin{align*}
p_i & =\operatorname{tr} \left(\sqrt{\rho}|i\rangle\langle i| \otimes I_m \sqrt{\rho}\right) \\
& =\operatorname{tr} \left(U^{\dagger}\left(A|i\rangle\langle i| \otimes I_m A^{\dagger}\right) U\right) \\
& =\operatorname{tr}\left(\eta_i U^{\dagger} \rho_i U\right)=\eta_i,
\end{align*}
and $\omega_i=U^{\dagger} \rho_i U$. Moreover,
\begin{align}
P_\textmd{S}^{\mathrm{opt}}\left(\left\{\rho_i, \eta_i\right\}_{i=1}^n\right)
&=\max _{\left\{M_i\right\}_{i=1}^n} \sum_i \eta_i \operatorname{tr}\left(M_i \rho_i\right) \nonumber\\
&=\max _{\left\{M_i\right\}_{i=1}^n} \sum_i \eta_i \operatorname{tr} \left(M_i U^{\dagger} \omega_i U \right) \nonumber\\
&=\max _{\left\{N_i\right\}_{i=1}^n} \sum_i \eta_i \operatorname{tr}\left(N_i \omega_i\right)\nonumber \\
&=P_\textmd{S}^{\mathrm{opt}}\left(\left\{\omega_i, \eta_i\right\}_{i=1}^n\right),\label{psopt}
\end{align}
where the last equality is due to the fact that if $\left\{M_i\right\}_{i=1}^n$ is a POVM on ${H}^{A}$, then $\left\{U M_i U^{\dagger}\right\}_{i=1}^n$ is a POVM on ${H}^A \otimes {H}^B$. Thus, we have the following result.

\textbf{Proposition 1}. For a set of quantum states $\rho_i$ ($i = 1, \cdots ,n$) of system $A$, based on parameterization process with respect to the POVM, the maximal success probability to identify $\{\rho_i,\eta_i\}$ is given by
\begin{align}
P^{\mathrm{opt}}_\mathrm{S}(\{\rho_i,\eta_i\})
=P_{\scriptscriptstyle\mathcal{M}}^{\scriptscriptstyle\Phi}(i|\varepsilon),
\end{align}
where $P_{\scriptscriptstyle\mathcal{M}}^{\scriptscriptstyle\Phi}(i|\varepsilon)$ is the probability distribution given by (\ref{gen}).

Proposition 1 can be easily proven by using (\ref{psopt}) and the definition of $P_{\scriptscriptstyle\mathcal{M}}^{\scriptscriptstyle\Phi}(i|\varepsilon)$,
\begin{align}
P^{\mathrm{opt}}_\mathrm{S}(\{\rho_i,\eta_i\})
=&P_\textmd{S}^{\mathrm{opt}}\left(\left\{\omega_i, \eta_i\right\}_{i=1}^n\right)\nonumber\\
=& \max _{\left\{N_i\right\}_{i=1}^n} \sum_i \eta_i \operatorname{tr}\left(N_i \omega_i\right) \nonumber\\
=&\max _{\left\{M^A_i\otimes I^B\right\}_{i=1}^n} \sum_i \eta_i\operatorname{tr}((U_\varepsilon\otimes I^{B}\omega_i U_\varepsilon^\dagger \otimes I^{B})(M_i^\dagger M_i\otimes I^{B}))\nonumber\\
=&\mathop{\mathrm{max}}_{\{M^A_i\otimes I^B\}}\sum_i\eta_iP_{i}^{\scriptscriptstyle\Phi}(i|\varepsilon)\nonumber
=P_{\scriptscriptstyle\mathcal{M}}^{\scriptscriptstyle\Phi}(i|\varepsilon),\label{psopt1}
\end{align}
where the corresponding PIO reads $N_i=M_iU_\varepsilon\otimes I^{B}$, and $\{M^A_i\}$ is the rank-1 POVM on the  subsystem $A$.

We also have $P^{\textmd{opt}}_\textmd{S}(\{\rho_i,\eta_i\})\geq P^{\textmd{opt (vN)}}_\textmd{S}(\{\rho_i,\eta_i\})$, and the maximal success probability to discriminate $\{\rho_i,\eta_i\}$  using von Neumann measurement on subsystem $A$ is represented by $P^{\textmd{opt
(vN)}}_\textmd{S}(\{\rho_i,\eta_i\})$.

Proposition 1 establishes an equivalence between the probability of a parameterized process based on POVMs and the maximum probability of successfully identifying  the quantum state $\{\rho_i,\eta_i\}$. One naturally finds a connection between the partial coherence measure based on the parameterized QFI and the maximum probability of successfully identifying the quantum state. We have the following conclusion.

\textbf{Proposition 2}.
For a set of quantum states $\rho_i$ ($i = 1, ..., n$) of system $A$, the QFI partial coherence and the maximal success probability have the following quantitative relationship
$$
\begin{aligned}
C_{\scriptscriptstyle Q}^{\scriptscriptstyle\varepsilon_0}(\rho^{AB})=\mathop{\max}\limits_{\Phi\in G}F_{\scriptscriptstyle Q}(\rho^{AB},\Phi,\varepsilon_0)=\mathop{\max}\limits_{\Phi\in G}\mathop{\max}\limits_{\mathcal{M}}F(P^{\textmd{opt}}_\textmd{S}(\{\rho_i,\eta_i\}),\varepsilon_0),\label{QFI}
\end{aligned}
$$
for the parametrization in (\ref{CQFI}), $C_{\scriptscriptstyle Q}^{\scriptscriptstyle\varepsilon_0}(\rho^{AB})$  is defined by QFI, and $P^{\textmd{opt}}_\textmd{S}(\{\rho_i,\eta_i\})$ is the maximum success probability of discriminating $\{\rho_i,\eta_i\}$  using the POVM on the system $A$.

In proposition 2, $C_{\scriptscriptstyle Q}^{\scriptscriptstyle\varepsilon_0}(\rho^{AB}) $ denotes the maximization over all possible parameterization processes $\Phi$ in $G$. This maximization reflects the maximum partial coherence that can be achieved for a given state $\rho^{AB}$ under a parameterized transformation. Proposition 2 shows the connection between the coherence of a quantum state and related the quantum state discrimination task. The measure \( C_{\scriptscriptstyle Q}^{\scriptscriptstyle\varepsilon_0}(\rho^{AB}) \) provides a tool in quantifying the partial coherence of quantum states relevant to the quantum information tasks.

\section{Conclusion and Discussion}\label{sect6}
\noindent

By accounting for the post-selective non-unitary parametrization process, we have established the partial coherence measures based on FI (and QFI).
We have shown that partial coherence is directly related to the precision of parameter estimation. This means that the degree of partial coherence reflects the ability of the quantum system to be able to be used for parameter estimation. Higher partial coherence indicates that the system has more information about the parameter, thus improving the estimation accuracy. By establishing a direct link between partial coherence and FI, our measures inherit the operational meaning of FI. This is because FI is not only used to quantify the information content of the quantum state about the parameter, but also determines the final accuracy limit of the parameter estimation.
Thus, our partial coherence measures capture the operational significance of FI in the context of quantum parameter estimation.

Furthermore, we have examined the differences between our partial coherence measure and the QFI under unitary parametrization. We have provided analytic expressions of the partial coherence measure for two-qubit states. Our partial coherence measures offer new insights into the significance of the non-unitary parametrization process. Moreover, we have provided operational significance of the partial coherence measures based on QFI in QSD. These results provide an interesting link between QSD and partial coherence theory.

Note that the complexity of calculating the FI and QFI increases with the scale of the systems. For larger systems, especially those with many degrees of freedom, direct computation of these measures can be computationally challenging. Our measures of partial coherence based on QFI can mitigate this issue in certain sense by taking into account the post-selective non-unitary parametrization process. Through the optimal estimation accuracy with non-unitary parametrization, our partial coherence measure naturally inherits the operational meaning of FI, as the optimal estimation accuracy is constrained by FI and asymptotically attained with maximum likelihood estimators. Our measures may be also experimentally implemented by using setups that involve non-linear optical processes or quantum interferometers. The partial coherence measures can be estimated from the interference patterns observed in these setups. Quantum metrology provides a natural platform for the implementation and estimation of these measures by manipulating quantum states through non-unitary operations and measuring the systems. The emerging field of quantum computing offers another avenue for implementing these measures. Quantum algorithms can be designed to simulate the non-unitary parametrization process, and quantum sensors can be used to detect changes in coherence, providing a direct measurement of partial coherence.

Our results would highlight further investigations on the relationship between the parameterized definition-based QFI partial coherence measures and quantum entanglement, quantum correlation (quantum discord) in the hope of further enriching the results of quantum resources.

\section*{ACKNOWLEDGEMENTS}
This work is supported by the National Natural Science Foundation of China (NSFC) under Grant Nos. 12075159 and 12171044; the specific research fund of the Innovation Platform for Academicians of Hainan Province.

\bigskip
\section*{APPENDIX}
\setcounter{equation}{0} \renewcommand%
\theequation{A\arabic{equation}}

\subsection{Proof of Theorem 1}
\textbf{Proof:} We prove that $C^{\scriptscriptstyle\varepsilon_0}_F(\rho^{AB})$ satisfy the conditions $(C1)$-$(C4)$.

$(C1)$ If $\sigma=\sum_ip_i\ket{i}\bra{i}\otimes\sigma_i$ is a partial incoherent state defined by (\ref{eq22}), so we have for each $\Phi$ and $l$,

\begin{eqnarray}
&&\Phi_l(\varepsilon)\sigma \Phi_l(\varepsilon)^\dagger \nonumber \\
=&& (K^A_l(\varepsilon) \otimes I^B) \sigma (K^A_l(\varepsilon) \otimes I^B)^\dagger \nonumber \\
=&& \left( \sum_n a_n^l(\varepsilon) |f_l(n)\rangle \langle n| \otimes I^B \right) \sigma \left( \sum_m a_m^{l*}(\varepsilon) |m\rangle \langle f_l(m)| \otimes I^B \right) \nonumber \\
=&& \sum_{nmi} a_n^l(\varepsilon) a_m^{l*}(\varepsilon) p_i |f_l(n)\rangle \langle n|i\rangle \langle i| m \rangle \langle f_l(m)| \otimes I^B \sigma_i I^B \nonumber \\
=&& \sum_n |a_n^l(\varepsilon)|^2 |f_l(n)\rangle \langle f_l(n)| \otimes \sigma_{n}\label{denominator1},
\end{eqnarray}
which is independent of $\varepsilon$ as $a_n^l(\varepsilon)=b_n^le^{ig_n^l(\varepsilon)}$.
Consequently, $P^{\scriptscriptstyle\Phi}(l|\varepsilon)$ is independent of $\varepsilon$, indicating that
\begin{align}
F(P^{\Phi},\varepsilon_0)=\sum_{l}\left[\left.\frac{\partial P^{\Phi}(l|\varepsilon)}{\partial\varepsilon}\right|_{\varepsilon_0}\right]^2\frac{1}
{P^{\Phi}(l|\varepsilon_0)}=0,
\end{align}
this results in $C^{\scriptscriptstyle\varepsilon_0}_F(\rho^{AB})=0$.
Conversely, if $C^{\scriptscriptstyle\varepsilon_0}_F(\rho^{AB})=0$, by the definition of $C^{\scriptscriptstyle\varepsilon_0}_F(\rho^{AB})$ we have  $\mathop{\max}\limits_{\Phi\in G} F(P^{\scriptscriptstyle\Phi},\varepsilon_0)=0$, which means that $F(P^{\scriptscriptstyle\Phi},\varepsilon_0)=0$ for all $\Phi\in G$.
From the definition of $F(P^{\scriptscriptstyle\Phi},\varepsilon_0)$, we obtain
\begin{align}
F(P^\Phi,\varepsilon_0)&=\sum_lP^\Phi(l|\varepsilon_0)\left[\left.\frac{\partial \ln P^\Phi(l|\varepsilon)}{\partial\varepsilon}\right|_{\varepsilon_0}\right]^2\nonumber\\
&=\sum_{l}\left[\left.\frac{\partial P^{\Phi}(l|\varepsilon)}{\partial\varepsilon}\right|_{\varepsilon_0}\right]^2\frac{1}
{P^{\Phi}(l|\varepsilon_0)}=0\label{denominator}.
\end{align}

From (\ref{denominator}) and the definition of $P^{\Phi}(l|\varepsilon)$, i.e., $P^{\scriptscriptstyle\Phi}(l|\varepsilon)=\operatorname{tr}(\Phi_l(\varepsilon)\rho^{AB} \Phi_l(\varepsilon)^\dagger)$, we have that $P^{\Phi}(l|\varepsilon) \neq 0$. Therefore, for any $l$ and $\varepsilon$, it must hold that $\frac{\partial \ln P^\Phi(l|\varepsilon)}{\partial\varepsilon}=0$. From (\ref{denominator1}), we see that $P^{\Phi}(l|\varepsilon)$ does not depend on the parameter $\varepsilon$ only if the quantum state is partially incoherent, namely, the final state $\Phi_l(\varepsilon)\rho^{AB} \Phi_l(\varepsilon)^\dagger$ will be the same as $\rho^{AB}$ in the parameterization process. Hence, one cannot obtain any information about the parameter $\varepsilon$ from the final measurements if the state is partially incoherent.

Furthermore, consider an $mm\times mm$-dimensional bipartite quantum state $\rho^{AB}$ with
non-zero partial coherence. $\rho^{AB}$ must have non-zero off-diagonal entries. Without loss of generality, assume $\rho^{AB}_{12}=|\rho^{AB}_{12}|e^{i\alpha}\neq 0$.
There exists a PIO  $\Phi(\varepsilon)=\{K^A_i(\varepsilon)\otimes I^B\}\in G$, where $\{K^A_i(\varepsilon)\}_i$ is the Kraus operator in the subsystem $A$, and $K^A_1(\varepsilon)=\frac{\sqrt{2}}{2}e^{i(\varepsilon+\gamma)}|1\rangle\langle 1|+\frac{\sqrt{2}}{2}|1\rangle\langle2|,\quad K^A_2(\varepsilon)=-\frac{\sqrt{2}}{2}e^{i(\varepsilon+\gamma)}|2\rangle\langle 1|+\frac{\sqrt{2}}{2}|2\rangle\langle 2|, \quad K^A_3(\varepsilon)=\sum_{n=3}^m|n\rangle\langle n|,$
with $\alpha+\varepsilon_0+\gamma\in[-\pi/2,0)\bigcup(0,\pi/2]$, such that
$P^{\scriptscriptstyle\Phi}(1|\varepsilon_0)=\operatorname{tr}(\Phi_1(\varepsilon_0)\rho^{AB} \Phi_1(\varepsilon_0)^\dagger)\neq0$ and $\partial_\varepsilon \operatorname{tr}(\Phi_1(\varepsilon)\rho^{AB} \Phi_1(\varepsilon)^\dagger )|_{\varepsilon_0}\neq 0$, which obviously shows that $C^{\scriptscriptstyle\varepsilon_0}_F(\rho^{AB})\neq 0$, namely, $C^{\scriptscriptstyle\varepsilon_0}_F(\rho^{AB})> 0$.

$(C3)$
Suppose a bipartite state $\rho^{AB}$ undergoes an arbitrary PIO $\tilde{\Phi}= \{\tilde{\Phi}_r\}_r  =\{K_r\otimes I^B\}_r$,
where $\tilde{\Phi}_r=K_r\otimes I^B=\sum_nb_n^r|f_r(n)\rangle\langle n|\otimes I^B$, $\{K_r\}_r$ is the Kraus operator in the subsystem $A$. The post-measurement ensemble $\{v_r,\rho^{AB}_r\}$ reads
$v_r=\operatorname{tr}(\tilde{\Phi}_r\rho^{AB} \tilde{\Phi}_r^\dagger)$ and $\rho^{AB}_r=\tilde{\Phi}_r\rho^{AB} \tilde{\Phi}_r^\dagger/v_r$.
Let $\Phi^{(r)}= \{{\Phi}^r_l\}_l=\{E_l^r(\varepsilon)\otimes I^B\}_l$ be the optimal PIO for $\rho^{AB}_r$ such that
$C^{\scriptscriptstyle\varepsilon_0}_F(\rho^{AB}_r)=F(P_r,\varepsilon_0)$, where $\Phi^{r}_{l}=E^r_l(\varepsilon)\otimes I^B
=\sum_nc_n^{rl}(\varepsilon)|g_{rl}(n)\rangle\langle n|\otimes I^B$ and
\begin{align*}
P_r(l|\varepsilon)
=&\operatorname{tr}(\Phi^{r}_{l}\rho^{AB}_r{\Phi^{r}_{l}}^\dagger)\\
=&\operatorname{tr}(E_l^r(\varepsilon)\otimes I^B\rho^{AB}_rE_l^{r}(\varepsilon)^\dagger\otimes I^B)\\
=&\operatorname{tr}(E_l^r(\varepsilon)K_r\otimes I^B\rho^{AB} K_r^\dagger E_l^{r}(\varepsilon)^\dagger\otimes I^B)/v_r\\
=&P(l,r|\varepsilon)/v_r.
\end{align*}

The $P(l,r|\varepsilon)$ above represents the probability distribution from
$\Phi'=\{E'_{lr}(\varepsilon)\otimes I^B\}_{lr}$ with
\begin{align*}
E'_{lr}(\varepsilon)\otimes I^B=E_l^r(\varepsilon)K_r\otimes I^B
=\sum_nb_n^rc_{f_r(n)}^{rl}(\varepsilon)|g_{rl}[f_r(n)]\rangle\langle n|\otimes I^B,
\end{align*}
which implies that $\Phi'\in G$. Therefore, one arrives at

\begin{align*}
\sum_rv_rC^{\scriptscriptstyle\varepsilon_0}_F(\rho^{AB}_r)
=&\sum_{r}v_rF(P_r,\varepsilon_0)\\
=&\sum_rv_r\sum_{l\in Y_r}\left[\left.\frac{\partial
P_r(l|\varepsilon)}{\partial\varepsilon}\right|_{\varepsilon_0}\right]^2\frac{1}{P_r(l|\varepsilon_0)}\\
=&\sum_rv_r\sum_{l\in Y_r}\left[\left.\frac{\partial
P(r,l|\varepsilon)}{\partial\varepsilon}\right|_{\varepsilon_0}\right]^2\frac{1}{P(r,l|\varepsilon_0)v_r}\\
=&\sum_r\sum_{l\in Y_r}\left[\left.\frac{\partial
P(r,l|\varepsilon)}{\partial\varepsilon}\right|_{\varepsilon_0}\right]^2\frac{1}{P(r,l|\varepsilon_0)}\\
=&F(P,\varepsilon_0)\leq C^{\scriptscriptstyle\varepsilon_0}_F(\rho^{AB}),
\end{align*}
where $Y_r$ indicates the region of $l$ in $P_r$, the last inequality is from that $\Phi'$ may not be the optimal one for $\rho^{AB}$.

$(C4)$ Let $\Phi=\{K_l(\varepsilon)\otimes I^B\}$ be the optimal PIO for
a bipartite state $\rho^{AB}=\sum_it_i\rho_i^{AB}$, in the sense that
$C^{\scriptscriptstyle\varepsilon_0}_F(\rho^{AB})=F(P,\varepsilon_0)$ with
$P(l|\varepsilon)=\operatorname{tr}(K_l(\varepsilon)\otimes I^B\rho^{AB} K_l(\varepsilon)^{\dagger}\otimes I^B)$. With respect to the state $\rho_i^{AB}$, denote
$P_i(l|\varepsilon)=
\operatorname{tr}(K_l(\varepsilon)\otimes I^B\rho_i^{AB} K_l(\varepsilon)^{\dagger}\otimes I^B)$. Then
\begin{align*}
\sum_it_iP_i(l|\varepsilon)
=&\sum_it_i\operatorname{tr}(K_l(\varepsilon)\otimes I^B\rho_i^{AB} K_l(\varepsilon)^{\dagger}\otimes I^B)\\
=&\operatorname{tr}(K_l(\varepsilon)\otimes I^B\rho^{AB} K_l(\varepsilon)^{\dagger}\otimes I^B)\\
=&P(l|\varepsilon).
\end{align*}

Still, as $\Phi$ might not be optimal for $\rho_i^{AB}$, i.e., $C^{\scriptscriptstyle\varepsilon_0}_F(\rho_i^{AB})\geq F(P_i,\varepsilon_0)$,
one gets
\begin{align*}
\sum_it_iC^{\scriptscriptstyle\varepsilon_0}_F(\rho_i^{AB})&\geq \sum_it_iF(P_i,\varepsilon_0)\\
\geq&F(\sum_it_iP_i,\varepsilon_0)=F(P,\varepsilon_0)\\
=&C^{\scriptscriptstyle\varepsilon_0}_F(\rho),
\end{align*}
where the convexity of FI is the reason of the second inequality.

Now, from the framework of partial coherence measurements, it is observed that partial coherence measures satisfying the conditions $(C3)$ and $(C4)$ satisfy the condition $(C2)$ too. This can be seen as follows:
\begin{align*}
C^A(\Phi^A(\rho^{AB}))=C^A(\sum_ip_i\rho^{AB}_i)\stackrel{(C4)}
{\leq}\sum_ip_iC^A(\rho^{AB}_i)
=\sum_ip_iC^A(p^{-1}_iK_i\rho^{AB}K^{\dagger}_i)
\stackrel{(C3)}{\leq} C^A(\rho^{AB}).
\end{align*}
Therefore, $C^{\scriptscriptstyle\varepsilon_0}_F(\rho^{AB})=\mathop{\max}\limits_{\Phi\in G}F(P^{\scriptscriptstyle\Phi},\varepsilon_0)$ is indeed a well-defined measure of partial coherence.

\subsection{Proof of Lemma 1}
\textbf{Proof:} Let $\Phi=\{K_l(\varepsilon)\otimes I^B\}\in G$. We have
\begin{align*}
K_l(\varepsilon)\otimes I^B
=&\sum_nb_n^le^{ig_n^l(\varepsilon)}|f_l(n)\rangle\langle n|\otimes I^B\\
=&\sum_nb_n^l|f_l(n)\rangle\langle n|\sum_me^{ig_m^l(\varepsilon)}|m\rangle\langle m|\otimes I^B\\
=&V_lU_l(\varepsilon)\otimes I^B,
\end{align*}
here $V_l=\sum_n$ $b_n^l|f_l(n)\rangle\langle n|$ and $U_l(\varepsilon)=\sum_m
e^{ig_m^l(\varepsilon)}|m\rangle\langle m|$.
Thus we have
\begin{align}
&(K_{l}(\varepsilon) \otimes I^B)^\dagger (K_{l}(\varepsilon)\otimes I^B)\nonumber\\
&=K_{l}(\varepsilon)^\dagger K_{l}(\varepsilon)\otimes I^BI^B \nonumber\\
&=U_{l}(\varepsilon)^\dagger V_{l}^\dagger V_{l}U_{l}(\varepsilon)\otimes I^B \nonumber\\
&=U_{l}(\varepsilon)^\dagger(\sum_i|\varphi_i^{l}\rangle\langle\varphi_i^{l}|)U_{l}(\varepsilon)\otimes I^B \nonumber\\
&=\sum_iU_{l}(\varepsilon)^\dagger|\varphi_i^{l}\rangle\langle\varphi_i^{l}|U_{l}(\varepsilon)\otimes I^B \nonumber\\
&=\sum_i|\phi_i^{l}(\varepsilon)\rangle\langle\phi_i^{l}(\varepsilon)| \otimes I^B =\sum_i\tilde{K}_{l,i}(\varepsilon)^\dagger\tilde{K}_{l,i}(\varepsilon)\otimes I^B,\label{77}
\end{align}
where $\sum_i|\varphi_i^{l}\rangle\langle\varphi_i^{l}|$ denotes the eigen-decomposition of $V_l^\dagger V_l$ (with the eigenvalue absorbed in $|\varphi_i^{l}\rangle$), $|\phi_i^{l}(\varepsilon)\rangle=U_{l}(\varepsilon)^\dagger|\varphi_i^{l}\rangle$ and
$\tilde{K}_{l,i}(\varepsilon)=|i\rangle\langle\phi_i^{l}(\varepsilon)|$. It is obvious that $\Phi'=
\{\tilde{K}_{l,i}(\varepsilon)\}_{li}\in G_1$.
From Cauchy-Schwarz inequality ($|\langle v|w\rangle|^2\leq\langle v|v\rangle\langle w|w\rangle$)  \cite{steele2004cauchy}, one obtains
\begin{align}
[\partial_{\varepsilon}P(l|\varepsilon)|_{\varepsilon_0}]^2\leq\sum_i\frac{[\partial_\varepsilon P_i(l|\varepsilon)|_{\varepsilon_0}]^2}{P_i(l|\varepsilon_0)}\sum_iP_i(l|\varepsilon_0),\label{pro}
\end{align}
where $P(l|\varepsilon)=\operatorname{tr}(\rho^{AB} \Phi_{l}^\dagger \Phi_{l})$ and $P_i(l|\varepsilon)=\operatorname{tr}(\rho^{AB} \tilde {\Phi}_{l,i}^\dagger \tilde {\Phi}_{l,i})$.
Consequently,
\begin{align}
\frac{[\partial_\varepsilon P(l|\varepsilon)|_{\varepsilon_0}]^2}{P(l|\varepsilon_0)}\leq
\sum_i\frac{[\partial_\varepsilon P_i(l|\varepsilon)|_{\varepsilon_0}]^2}{P_i(l|\varepsilon_0)},\label{resul}
\end{align}
here the inequality holds for all values of $l$. Therefore, $F(P^{\scriptscriptstyle\Phi},\varepsilon_0)\leq F(P^{\scriptscriptstyle\Phi'},\varepsilon_0)$.

\subsection{Proof of  Lemma 2}

\textbf{Proof:} Assume that $\tilde{\Phi}$ and $\mathcal{M}$, respectively, represent the optimal measurement and parametrization for the optimal $F_{\scriptscriptstyle Q}$. From (\ref{gen}) we have $P_{\scriptscriptstyle\mathcal{M}}^{\scriptscriptstyle\tilde{\Phi}}(l|\varepsilon)
$$=\operatorname{tr}((M^A_l\otimes I^B)\tilde{\Phi}_\varepsilon(\rho^{AB}))$ $=\operatorname{tr}((\sum_i|\varphi_i^l\rangle\langle\varphi_i^l|\otimes I^B)\tilde{\Phi}_\varepsilon(\rho^{AB}))$ $=\sum_iP_i(l|\varepsilon)$, where $\sum_i|\varphi_i^l\rangle\langle\varphi_i^l|$
represents the eigen-decomposition of $M_l^A$. In particular, let $P_i(l|\varepsilon)=\operatorname{tr}(\left\langle\varphi_i^l\right\vert\otimes I^B(\tilde{\Phi}_\varepsilon(\rho^{AB}))\left\vert\varphi_i^l\right\rangle\otimes I^B)$ can be expressed as
\begin{align*}
P_i(l|\varepsilon)=&\operatorname{tr}(\left\vert i\right\rangle\left\langle\varphi_i^l\right\vert\otimes I^B(\tilde{\Phi}_\varepsilon(\rho^{AB}))\left\vert\varphi_i^l\right\rangle\left\langle i\right\vert\otimes I^B)\cr\cr
=&\sum_{ynm}\operatorname{tr}(\left\vert i\right\rangle\left\langle\varphi_i^l\right\vert\otimes I^B
a_n^y(\varepsilon)|f_y(n)\rangle\langle n| \otimes I^B\rho^{AB} a_m^{y*}(\varepsilon)
|m\rangle\langle f_y(m)|\otimes I^B \left\vert\varphi_i^l\right\rangle\left\langle i\right\vert\otimes I^B)\\
=&\sum_{ynm}\operatorname{tr}(a_{n}^{ily}(\varepsilon)\left\vert i\right\rangle\langle n|\otimes I^B\rho^{AB} |m\rangle\left\langle i\right\vert a_{m}^{ily*}(\varepsilon)\otimes I^B)\\
=&\sum_y\operatorname{tr}(K_{ily}(\varepsilon) \otimes I^B\rho^{AB} K_{ily}(\varepsilon)^\dagger\otimes I^B)=\sum_yP(ily|\varepsilon),
\end{align*}
here $a_{n}^{ily}(\varepsilon)=\left\langle\varphi_i^l\right\vert a_n^y(\varepsilon)|f_y(n)\rangle$
and $K_{ily}(\varepsilon)=\sum_{n}a_{n}^{ily}(\varepsilon)\left\vert i\right\rangle\langle n|$.

It is obvious that $\Phi'_\varepsilon=\{K_{ily}(\varepsilon)\otimes I^B\}\in G$. Then
\begin{align*}
&\mathop{\max}\limits_{\Phi}F_{\scriptscriptstyle Q}(\rho^{AB},\Phi,\varepsilon_0)=F(P_{\scriptscriptstyle\mathcal{M}}^{\scriptscriptstyle\tilde{\Phi}},\varepsilon_0)\nonumber\\
&=\sum_l\frac{[\partial_\varepsilon P_{\scriptscriptstyle\mathcal{M}}^{\scriptscriptstyle\tilde{\Phi}}(l|\varepsilon)|_{\varepsilon_0}]^2}{P_{\scriptscriptstyle\mathcal{M}}^{\scriptscriptstyle\tilde{\Phi}}(l|\varepsilon_0)}=
\sum_l\frac{[\sum_i\partial_\varepsilon P_i(l|\varepsilon)|_{\varepsilon_0}]^2}{\sum_iP_i(l|\varepsilon_0)}\nonumber\\
&\leq\sum_{il}\frac{[\partial_\varepsilon P_i(l|\varepsilon)|_{\varepsilon_0}]^2}{P_i(l|\varepsilon_0)}
=\sum_{il}\frac{[\sum_y\partial_\varepsilon P(ily|\varepsilon)|_{\varepsilon_0}]^2}{\sum_yP(ily|\varepsilon_0)}\nonumber\\
&\leq\sum_{ily}\frac{[\partial_\varepsilon P(ily|\varepsilon)|_{\varepsilon_0}]^2}{P(ily|\varepsilon_0)}=F(P,\varepsilon_0)\leq\mathop{\max}\limits_{\Phi}
F(P^{\scriptscriptstyle\Phi},\varepsilon_0),
\end{align*}
where $P$ is the distribution from $\Phi'_\varepsilon$, the first two inequalities are derived from Cauchy-Schwarz inequality. Similar to the derivations of Eq. (\ref{pro}) and (\ref{resul}), from
\begin{align*}
\left[\sum_i\partial_\varepsilon P_i(l|\varepsilon)|_{\varepsilon_0}\right]^2\leq \sum_i\frac{[\partial_\varepsilon P_i(l|\varepsilon)|_{\varepsilon_0}]^2}{P_i(l|\varepsilon_0)}\sum_iP_i(l|\varepsilon_0),
\end{align*}
we obtain the first inequality,
and from
\begin{align*}
\left[\sum_y\partial_\varepsilon P(ily|\varepsilon)|_{\varepsilon_0}\right]^2\leq\sum_y\frac{[\partial_\varepsilon P(ily|\varepsilon)|_{\varepsilon_0}]^2}{P(ily|\varepsilon_0)}\sum_yP(ily|\varepsilon_0),
\end{align*}
we get the second inequality.

\subsection{Proof of Theorem 2}
\textbf{Proof:}
By the  Lemma 1, $C^{\scriptscriptstyle\varepsilon_0}_F(\rho^{AB})$ can be written as
$C^{\scriptscriptstyle\varepsilon_0}_F(\rho^{AB})=\mathop{\max}\limits_{\Phi\in G_1}
F(P^{\scriptscriptstyle\Phi},\varepsilon_0)$. Suppose $\Phi=\{K_j(\varepsilon)\otimes I^B\}$ is the optimal operation in $G_1$ and $\{K_j(\varepsilon)\}_j$ is the Kraus on subsystem $A$ such that
$C^{\scriptscriptstyle\varepsilon_0}_F(\rho^{AB})=F(P^{\scriptscriptstyle\Phi},\varepsilon_0)$,
where $P^{\scriptscriptstyle\Phi}(j|\varepsilon)=\operatorname{tr}(K_j(\varepsilon)\otimes I^B\rho^{AB} K_j(\varepsilon)^\dagger\otimes I^B)$ and $Rank[K_j(\varepsilon)^\dagger K_j(\varepsilon)]=1$.
Without lost of generality, $K_j(\varepsilon)$ can be written as $K_j(\varepsilon)=|j\rangle\langle\phi_j(\varepsilon)|$.
Denote $P^{\scriptscriptstyle\Phi}_{\scriptscriptstyle\mathcal{P}}(j|\varepsilon)
=\operatorname{tr}(|j\rangle\langle j|\otimes I^B\Phi_\varepsilon(\rho^{AB})|j\rangle\langle j|\otimes I^B)$, where $\mathcal{P}=\{|j\rangle\langle j|\otimes I^B\}_j$ represents the projective measurements on the parameterized state.
Since
\begin{align*}
&P^{\scriptscriptstyle\Phi}(j|\varepsilon)=\operatorname{tr}(K_j\otimes I^B\rho^{AB} K_j^\dagger\otimes I^B)\\
=&\operatorname{tr}(|j\rangle\langle j|\otimes I^B(\sum_{j'}K_{j'}\otimes I^B\rho^{AB} K_{j'}^\dagger\otimes I^B)|j\rangle\langle j|\otimes I^B)\\
=&\operatorname{tr}(|j\rangle\langle j|\otimes I^B\Phi_\varepsilon(\rho^{AB})|j\rangle\langle j|\otimes I^B)=P^{\scriptscriptstyle\Phi}_{\scriptscriptstyle\mathcal{P}}(j|\varepsilon),
\end{align*}
we have
\begin{align*}
&C^{\scriptscriptstyle\varepsilon_0}_F(\rho^{AB})=F(P^{\scriptscriptstyle\Phi},\varepsilon_0)=F(P^{\scriptscriptstyle\Phi}_{\scriptscriptstyle\mathcal{P}},\varepsilon_0)
\leq \mathop{\max}\limits_{\mathcal{M}}F(P_{\scriptscriptstyle\mathcal{M}}^{\scriptscriptstyle\Phi},\varepsilon_0) \\
=&F_{\scriptscriptstyle Q}(\rho^{AB},\Phi,\varepsilon_0)\leq\mathop{\max}\limits_{\Phi\in G}F_{\scriptscriptstyle Q}(\rho^{AB},\Phi,\varepsilon_0)=C_{\scriptscriptstyle Q}^{\scriptscriptstyle\varepsilon_0}(\rho^{AB}).\label{project}
\end{align*}

Conversely, from Lemma 2 we immediately obtain $C^{\scriptscriptstyle\varepsilon_0}_F(\rho^{AB})\geq C_{\scriptscriptstyle Q}^{\scriptscriptstyle\varepsilon_0}(\rho^{AB})$. Thus one gets $C_{\scriptscriptstyle Q}^{\scriptscriptstyle\varepsilon_0}(\rho^{AB})=C^{\scriptscriptstyle\varepsilon_0}_F(\rho^{AB})$, which finishes the proof.

\subsection{Proof of Theorem 3}

\textbf{Proof:} For a $2\otimes2$ state $\rho^{AB}$, consider PIO $\{K_l\otimes I^{B}\}\in G$ with
\begin{align*}
K_l(\varepsilon)=b_1'^le^{i g_1'^l(\varepsilon)}|f_l(1)\rangle\langle 1|+b_2'^le^{ig_2'^l(\varepsilon)}|f_l(2)\rangle\langle 2|,
\end{align*}
where $b_1'^l$ or $b_2'^l$ may be zero. The Kraus operator can be written as
\begin{align*}
K_l(\varepsilon)=b_1^le^{ig_1^l(\varepsilon)}|f_l(1)\rangle\langle 1|+b_2^le^{i g_2^l(\varepsilon)}|f_l(2)\rangle\langle 2|,
\end{align*}
where $b_j^l=b_j'^le^{ig_j'^l(\varepsilon_0)}$ and $g_j^l(\varepsilon)=g_j'^l(\varepsilon)-g_j'^l(\varepsilon_0)$ for $j=1,2$.
According to Lemma 1 and its proof,
the optimal PIO can be rank-1 with the form $\{\left\vert i\right\rangle\left\langle \varphi_i^l(\varepsilon)\right\vert\}$, which means $f_l(1)=f_l(2)$ for any $l$. Then we have
\begin{align*}
P(l|\varepsilon)
=&\operatorname{tr}
((K_{l}(\varepsilon) \otimes I^B) \rho (K_{l}(\varepsilon)\otimes I^B)^\dagger)\nonumber\\
=&2(|b_1^l|^2\rho_{11}+|b_1^l|^2\rho_{22} +|b_2^l|^2\rho_{33}+|b_2^l|^2\rho_{44}\\
&+\rho_{13}b_1^lb_2^{l*}e^{i[g_1^l(\varepsilon)-g_2^l(\varepsilon)]}
+\rho_{31}b_1^{l*}b_2^{l}e^{-i[g_1^l(\varepsilon)-g_2^l(\varepsilon)]}
+\rho_{24}b_1^lb_2^{l*}e^{i[g_1^l(\varepsilon)-g_2^l(\varepsilon)]}
+\rho_{42}b_1^{l*}b_2^{l}e^{-i[g_1^l(\varepsilon)-g_2^l(\varepsilon)]}).
\end{align*}
Therefore

\begin{align*}
F(P,\varepsilon_0)=&\sum_l\frac{[4\operatorname{Im}(\rho_{13}b_1^lb_2^{l*})]^2[\partial_\varepsilon g_1^l(\varepsilon)|_{\varepsilon_0}-\partial_\varepsilon g_2^l(\varepsilon)|_{\varepsilon_0}]^2+[4\operatorname{Im}(\rho_{24}b_1^lb_2^{l*})]^2[\partial_\varepsilon g_1^l(\varepsilon)|_{\varepsilon_0}-\partial_\varepsilon g_2^l(\varepsilon)|_{\varepsilon_0}]^2}{2|b_1^l|^2\rho_{11}+2|b_1^l|^2\rho_{22}+2|b_2^l|^2\rho_{33}+2|b_2^l|^2\rho_{44}
+4\operatorname{Re}(\rho_{13}b_1^lb_2^{l*})+4\operatorname{Re}(\rho_{24}b_1^lb_2^{l*})} \nonumber\\
\leq&\sum_l\frac{[4\operatorname{Im}(\rho_{13}b_1^lb_2^{l*})]^2+[4\operatorname{Im}(\rho_{24}b_1^lb_2^{l*})]^2}{2|b_1^l|^2\rho_{11}+2|b_1^l|^2\rho_{22}+2|b_2^l|^2\rho_{33}+2
|b_2^l|^2\rho_{44}+4\operatorname{Re}(\rho_{13}b_1^lb_2^{l*})+4\operatorname{Re}(\rho_{24}b_1^lb_2^{l*})},
\end{align*}
where the inequality may be saturated by $g_1^l(\varepsilon)=\varepsilon$ and $g_2^l(\varepsilon)=0$. The corresponding PIO reads $K_l(\varepsilon)\otimes I^{B}=E_lU_\varepsilon\otimes I^{B}$, where
$E_l=a_1^l|f_l(1)\rangle\langle 1|+a_2^l|f_l(2)\rangle\langle 2|$ and
$U_\varepsilon=e^{i\varepsilon}|1\rangle\langle 1|+|2\rangle\langle 2|$ with $f_l(1)=f_l(2)$ and $\{E_l\otimes I^{B}\}\in G_1$. The probability distribution can be expressed in this way
\begin{align*}
P^{\scriptscriptstyle\Phi}(l|\varepsilon)
&=\operatorname{tr}(K_l(\varepsilon)\otimes I^{B}\rho K_l(\varepsilon)^\dagger\otimes I^{B})\\
&=\operatorname{tr}(E_lU_\varepsilon\otimes I^{B} \rho U_\varepsilon^\dagger E_l^\dagger\otimes I^{B})\nonumber\\
&=\operatorname{tr}((U_\varepsilon\otimes I^{B}\rho U_\varepsilon^\dagger \otimes I^{B})(E_l^\dagger\otimes I^{B} E_l\otimes I^{B}))\nonumber\\
&=\operatorname{tr}((U_\varepsilon\otimes I^{B}\rho U_\varepsilon^\dagger \otimes I^{B})(E_l^\dagger E_l\otimes I^{B}))\\
&=P^{\scriptscriptstyle\Phi}_{\scriptscriptstyle\mathcal{M}}(l|\varepsilon).
\end{align*}
One can generate the probability distribution $P_{\scriptscriptstyle\mathcal{M}}$ using a unitary parametrization $U_\varepsilon$, and then performing a rank-1 POVM measurement $\mathcal{M}=\{E_l^\dagger E_l\otimes I^{B}\}$ on subsystem $A$.
From the above optimal PIO, one arrives at
$C^{\varepsilon_0}_F(\rho)=\mathop{\max}\limits_{\Phi\in G_1}
F(P^{\scriptscriptstyle\Phi},\varepsilon_0)=\mathop{\max}\limits_{\mathcal{M}}
F(P^{\scriptscriptstyle\Phi}_{\scriptscriptstyle\mathcal{M}},\varepsilon_0)=F_{\scriptscriptstyle Q}(\rho,U_\varepsilon,\varepsilon_0)$, this ends the proof.

\subsection{Derivations in Example 1}
Consider a bipartite state $\rho_1=|\phi_A\rangle\otimes|\phi_B\rangle$ with $|\phi\rangle_A=(\frac{1}{\sqrt{2}},\frac{1}{2},\frac{1}{2})^T$ and $|\phi\rangle_B=(\frac{1}{\sqrt{2}},0,\frac{1}{\sqrt{2}})^T$, and the parametrization $\Phi=\{K_l(\varepsilon)\otimes I^{B}\}$ given by
\begin{align*}
K_l(\varepsilon)\otimes I^{B}=&(b_1^le^{ig_1^l\varepsilon}|f_l(1)\rangle\langle 1|
+b_2^le^{ig_2^l\varepsilon}|f_l(1)\rangle\langle 2|\\
+&b_3^le^{ig_3^l\varepsilon}|f_l(1)\rangle\langle 3|)\otimes I^{B},
\end{align*}
where, by denoting $B^l=[b_1^l,b_2^l,b_3^l]$,
$B^1=[0,\sqrt{\frac{2}{15}},\sqrt{\frac{1}{5}}]$, $B^2=[0,\sqrt{\frac{2}{15}}e^{-i2\pi/3},\sqrt{\frac{1}{5}}e^{i2\pi/3}]$,
$B^3=[0,\sqrt{\frac{2}{15}}e^{-i4\pi/3},\sqrt{\frac{1}{5}}e^{i4\pi/3}]$,
$B^4=[\sqrt{\frac{2}{15}},\sqrt{\frac{1}{5}},0]$, $B^5=[\sqrt{\frac{2}{15}},\sqrt{\frac{1}{5}}e^{i2\pi/3},0]$, $B^6=[\sqrt{\frac{2}{15}},\sqrt{\frac{1}{5}}e^{i4\pi/3},0]$, $B^7=[\sqrt{\frac{1}{5}}, 0, \sqrt{\frac{2}{15}}]$, $B^8=[\sqrt{\frac{1}{5}}, 0, \sqrt{\frac{2}{15}}e^{i2\pi/3}]$,
$B^9=[\sqrt{\frac{1}{5}}, 0, \sqrt{\frac{2}{15}}e^{i4\pi/3}]$, and $g_1^l=0$, $g_2^l=1$, $g_3^l=0$ for $l=1,2,3$, $g_1^l=1$, $g_2^l=0$, $g_3^l=0$ for $l=4,\cdots,9$.
The probability distribution is given by
\begin{align*}
P(l|0)=&\operatorname{tr}(K_l(\varepsilon)\otimes I^{B}\rho K_l(\varepsilon)^\dagger\otimes I^{B})\\
=&\operatorname{tr}(K_l(\varepsilon)\otimes I^{B}(|\phi\rangle_A\langle\phi| \otimes |\phi\rangle_B \langle\phi|) K_l(\varepsilon)^\dagger\otimes I^{B})\nonumber\\
=&\operatorname{tr}(K_l(\varepsilon)|\phi\rangle_A\langle\phi|K_l(\varepsilon)^\dagger \otimes I^{B} |\phi\rangle_B \langle\phi|I^{B})\nonumber\\
=&\operatorname{tr}(K_l(\varepsilon)|\phi\rangle_A\langle\phi|K_l(\varepsilon)^\dagger \otimes  |\phi\rangle_B \langle\phi|)\nonumber\\
=&\operatorname{tr}(K_l(\varepsilon)|\phi\rangle_A\langle\phi|K_l(\varepsilon)^\dagger \operatorname{tr} (|\phi\rangle_B \langle\phi|)\nonumber\\
=&\phi_{11}|b_1^l|^2+\phi_{22}|b_2^l|^2+\phi_{33}|b_3^l|^2
+2\operatorname{Re}[\phi_{12}b_1^lb_2^{l*}+\phi_{23}b_2^lb_3^{l*}+\phi_{31}b_3^lb_1^{l*}],
\end{align*}
and
\begin{align*}
\partial_\varepsilon P(l|\varepsilon)|_0=&2\operatorname{Im}[\phi_{12}b_1^lb_2^{l*}(g_1^l-g_2^l)
+\phi_{23}b_2^lb_3^{l*}(g_2^l-g_3^l)+\phi_{31}b_3^lb_1^{l*}(g_3^l-g_1^l)].
\end{align*}
As a result, we have the corresponding FI,
\begin{align*}
F(P^\Phi,0)=\sum_l\frac{[\partial_\varepsilon P(l|\varepsilon)|_0]^2}{P(l|0)}=1.07.
\end{align*}
We obtain $C^0_F(\rho_1)\geq F(P^\Phi,0)$ from the definition (\ref{d22}).

To compare our measure with QFI subject to the optimal unitary parametrization in $G$, let us calculate $\mathop{\max}\limits_{U_\varepsilon\in G}F_{\scriptscriptstyle Q}(\rho,U_\varepsilon,0)$, where $U_\varepsilon$ is the unitary operator, $U_\varepsilon=\sum_ne^{ig_n(\varepsilon)}|n\rangle\langle n|\otimes I^{B}$ with $\partial_\varepsilon g_n(\varepsilon)\in[0,1]$. When the eigenvalues of the parameterized state $U_\varepsilon\rho U_\varepsilon^\dagger$ are parameter-independent, the local quantum Fisher information  can be written as \cite{WZJ,VSL},
$$
F_{\scriptscriptstyle Q}(\rho,U_\varepsilon,\varepsilon_0)=\operatorname{tr}\left(\rho H^2\right)-\sum_{i \neq j} \frac{2 P_i P_j}{P_i+P_j}|\langle\varphi_i|H|\varphi_j\rangle|^2,
$$
where the eigenvalues and eigenvectors of $U_\varepsilon\rho U_\varepsilon^\dagger$ are denoted by the $\{P_i\}$ and $\{|\varphi_i\rangle\}$, respectively, and terms with $P_i=P_j=0$ are excluded from the summation.

Assume that the Hamiltonian $H=H_A \otimes I_B$ governs the dynamics of the first system.
Consider the following Hamiltonian $H_\varepsilon$,
$$H_\varepsilon=\sum_n\partial_\varepsilon g_n(\varepsilon)|n\rangle\langle n|\otimes I^{B}.$$
For a bipartite pure state $\rho_1=|\varphi\rangle\langle\varphi|$, let $\{|\varphi_i\rangle\}$ be the basis vectors with $|\varphi\rangle=|\varphi_1\rangle$. Then correspondingly $P_1=1$ and the residual eigenvalues $P_i$ ($i\neq1$) are zero, and the eigenvectors of $U_\varepsilon\rho U_\varepsilon^\dagger$ are $\{U_\varepsilon|\varphi_i\rangle\}$. We have
$$F_{\scriptscriptstyle Q}(|\varphi\rangle,U_\varepsilon,\varepsilon_0)
=4\langle\varphi|H^2_{\varepsilon_0}|\varphi\rangle-4\langle\varphi|H_{\varepsilon_0}|\varphi\rangle^2.$$

As long as $|\varphi\rangle=|\varphi_1\rangle$, the conclusion is independent of the choice of $|\varphi_i\rangle$, and the optimal QFI $\mathop{\max}\limits_{U_\varepsilon\in G}F_{\scriptscriptstyle Q}(|\varphi\rangle,U_\varepsilon,0)$ can be computed as
\begin{align*}
\mathop{\max}\limits_{U_\varepsilon\in G}F_{\scriptscriptstyle Q}(|\varphi\rangle,U_\varepsilon,0)=&\mathop{\max}\limits_{H\in S}4\langle\varphi|H^2_{\varepsilon_0}|\varphi\rangle-4\langle\varphi|H_{\varepsilon_0}|\varphi\rangle^2=1.00,
\end{align*}
where $|\varphi\rangle=|\phi_A\rangle\otimes|\phi_B\rangle$ is the $3\otimes3$  bipartite state $\rho$ given above, $S$ is the set of operator $H=(g_1|1\rangle\langle1|+g_2|2\rangle\langle2|+g_3|3\rangle\langle3|)\otimes I^{B}$ ($g_i\in[0,1]$). Thus $C^0_F(\rho_1)>\mathop{\max}\limits_{U_\varepsilon\in G}F(|\varphi\rangle,U_\varepsilon,0)$, which implies that the FI with unitary parameterization is not the same as $C^{\scriptscriptstyle\varepsilon_0}_F$.

\subsection{ Region of  $\partial_\theta g(\varepsilon)$}

We first consider the case that $\frac{\partial g_n^l(\varepsilon)}{\partial \varepsilon}$ is finite and $\max _{n, l}\left|\frac{\partial g_n^l(\varepsilon)}{\partial \varepsilon}\right| \leqslant k$ with finite $k$. Let $\tilde{C}_{F_k}^{\varepsilon_0}$ denote a function analogous to $C^{\scriptscriptstyle\varepsilon_0}_F$ from Theorem 1, differing only by the condition $\max _{n, l}\left|\frac{\partial g_n^l(\varepsilon)}{\partial \varepsilon}\right| \leqslant k$. Denote $G^{(k)}$ the collection of the corresponding channels, namely,
\begin{align}
\tilde{C}_{F_k}^{\varepsilon_0}(\rho)=\max _{\Phi \in G^{(k)}} F\left(\tilde{P}^{\Phi}, \varepsilon_0\right),
\end{align}
where
\begin{align}\label{AQ6}
\tilde{P}^{\scriptscriptstyle\Phi}(l|\varepsilon)=\operatorname{tr}(\tilde{\Phi}_l(\varepsilon)\rho^{AB} \tilde{\Phi}_l(\varepsilon)^\dagger) =\operatorname{tr}((\tilde{K}^A_l(\varepsilon)\otimes I^B)\rho^{AB} (\tilde{K}^A_l(\varepsilon)\otimes I^B)^\dagger),
\end{align}
with $\tilde{\Phi}_l(\varepsilon)=\tilde{K}^A_l(\varepsilon)\otimes I^B=\sum_nc_n^le^{ig_n^l(\varepsilon)}|f_l(n)\rangle\langle n|\otimes I^B$ and $\max _{n, x}\left|\frac{\partial g_n^l(\varepsilon)}{\partial \varepsilon}\right| \leqslant k$. We observe that $\tilde{C}_{F_k}^{\varepsilon_0}$ bears a relation to the previous coherence measure. We first prove the following lemma.

Lemma 3. The function $\tilde{C}_{F_k}^{\varepsilon_0}$ satisfies that
$$
C_F^{\omega_0}(\rho)=\frac{4}{9 k^2} \tilde{C}_{F_k}^{\varepsilon_0}(\rho),
$$
where $\omega_0=\frac{3}{2}k \varepsilon_0$.

Proof. Suppose $\left\{{K}^A_l(\varepsilon)\otimes I^B\right\}$ are Kraus operators of the channel in $G^{(\frac{2}{3})}$, namely,
$$
{K}^A_l(\varepsilon)\otimes I^B=\sum_nc_n^le^{iu_n^l(\varepsilon)}|f_l(n)\rangle\langle n|\otimes I^B,
$$
where $\left|\partial_\omega u_n^l\right| \leqslant \frac{2}{3}$. Define
$$
\hat{K}^A(\omega)\otimes I^B=e^{i \frac{\omega}{3}} K^A_l(\omega)\otimes I^B=\sum_nc_n^le^{iv_n^l(\varepsilon)}|f_l(n)\rangle\langle n|\otimes I^B,
$$
where $v_n^l(\omega)=u_n^l(\omega)+\frac{\omega}{3}$ and thus $\partial_\omega v_n^l \in[0,1]$.
We have
\begin{align}
\hat{K}^A(\omega)\otimes I^B\rho (\hat{K}^A(\omega)\otimes I^B)^{\dagger} & =e^{i 2\omega / 3} K^A_l(\varepsilon)\otimes I^B \rho e^{-i 2\omega /3}(K^A_l(\varepsilon)\otimes I^B)^{\dagger}\nonumber \\
&= K^A_l(\varepsilon)\otimes I^B \rho (K^A_l(\varepsilon)\otimes I^B)^{\dagger},\nonumber
\end{align}
from which we get
\begin{align}
C_F^{\omega_0}(\rho)=\tilde{C}_{F_\frac{2}{3}}^{\omega_0}(\rho).\label{AQ7}
\end{align}

Concerning the $\left\{\tilde{K}^A_l(\varepsilon)\otimes I^B\right\}$  in Eq.(\ref{AQ6}), 
we obtain
$$
\tilde{P}(l \mid \varepsilon)=\sum_{nn^{\prime}}c_n^l c_{n^{\prime}}^{l *}e^{i\left[g_n^x(\varepsilon)-g_{n^{\prime}}^x(\varepsilon)\right]}|f_l(n)\rangle\langle n|\otimes I^B\rho | n^{\prime} \rangle\langle f_l(n^{\prime})| \otimes I^B,
$$
and
$$
\begin{aligned}
\left.\partial_\varepsilon \tilde{P}(l \mid \varepsilon)\right|_{\varepsilon_0}
=&\sum_{nn^{\prime}}\left\{c_n^lc_{n^{\prime}}^{l *}e^{i\left[g_n^x(\varepsilon)-g_{n^{\prime}}^x(\varepsilon)\right]}|f_l(n)\rangle\langle n|\otimes I^B\rho | n^{\prime} \rangle\langle f_l(n^{\prime})| \otimes I^B \quad \times i\left[\left.\partial_\varepsilon g_n^x(\varepsilon)\right|_{\varepsilon_0}-\left.\partial_\varepsilon g_{n^{\prime}}^x(\varepsilon)\right|_{\varepsilon_0}\right]\right\}.
\end{aligned}
$$

Set $\omega=\frac{3}{2}k\varepsilon$. Then
$$
\tilde{P}(l \mid \varepsilon)= \sum_{n, n^{\prime}}c_n^lc_{n^{\prime}}^{l *}\exp \left\{i\left[g_n^x\left(\frac{2\omega}{3 k}\right)-g_{n^{\prime}}^x\left(\frac{2\omega}{3 k}\right)\right]\right\}|f_l(n)\rangle\langle n|\otimes I^B\rho | n^{\prime} \rangle\langle f_l(n^{\prime})| \otimes I^B.
$$
Thus $\tilde{P}(l \mid \varepsilon)$ can be expressed in terms of $P(l \mid \omega)$. Define $h_n^x(\omega)=g_n^x\left(\frac{2\omega}{3 k}\right)$. Then $\partial_\omega h_n^x(\omega)=\frac{2\partial_\varepsilon g_n^x(\varepsilon)}{3 k}$ and thus $\left|\partial_\omega h_n^x(\omega)\right| \leqslant$ $\frac{2}{3}$. In addition,
$$
\begin{aligned}
&\left.\partial_\omega P(l \mid \omega)\right|_{\omega_0} \\
&= \sum_{n, n^{\prime}}\left(c_n^lc_{n^{\prime}}^{l *}\exp \left\{i\left[g_n^x\left(\frac{2\omega}{3 k}\right)-g_{n^{\prime}}^x\left(\frac{2\omega}{3 k}\right)\right]\right\}|f_l(n)\rangle\langle n|\otimes I^B\rho | n^{\prime} \rangle\langle f_l(n^{\prime})| \otimes I^B \quad \times i\left[\left.\partial_\omega g_n^x\left(\frac{2\omega}{3k}\right)\right|_{\omega_0}-\left.\partial_\omega g_{n^{\prime}}^x\left(\frac{2\omega}{3 k}\right)\right|_{\omega_0}\right]\right) \\
&= \frac{2}{3k} \sum_{nn^{\prime}}\left\{c_n^lc_{n^{\prime}}^{l *}e^{i\left[g_n^x(\varepsilon)-g_{n^{\prime}}^x(\varepsilon)\right]}|f_l(n)\rangle\langle n|\otimes I^B\rho | n^{\prime} \rangle\langle f_l(n^{\prime})| \otimes I^B \quad \times i\left[\left.\partial_\varepsilon g_n^x(\varepsilon)\right|_{\varepsilon_0}-\left.\partial_\varepsilon g_{n^{\prime}}^x(\varepsilon)\right|_{\varepsilon_0}\right]\right\} \\
&=\left.\frac{2}{3 k} \partial_\varepsilon \tilde{P}(l\mid \varepsilon)\right|_{\varepsilon_0},
\end{aligned}
$$
where $\omega_0=\frac{3 }{2k}\varepsilon_0$. Thus
$$
\begin{aligned}
F\left(P, \omega_0\right) & =\sum_x \frac{\left[\left.\partial_\omega P(x \mid \omega)\right|_{\omega_0}\right]^2}{P\left(x \mid \omega_0\right)} \\
& =\sum_x \frac{4\left[\left.\partial_\varepsilon \tilde{P}(x \mid \varepsilon)\right|_{\varepsilon_0}\right]^2}{9 k^2 \tilde{P}\left(x \mid \varepsilon_0\right)}=\frac{4F\left(\tilde{P}, \varepsilon_0\right)}{9 k^2}.
\end{aligned}
$$
Combining this with Eq.(\ref{AQ7}), we have
$$
C_F^{\omega_0}(\rho)=\frac{4}{9 k^2} \tilde{C}_{F_k}^{\varepsilon_0}(\rho).
$$

The above derivation indicates that should $\partial_\varepsilon g_n^x(\varepsilon)$ be finite, examining the scenario where $\partial_\varepsilon g_n^x(\varepsilon) \in[0,1]$ suffices to encompass all other possible general scenarios. If $\partial_\varepsilon g_n^x(\varepsilon)$ is infinite, then the FI and $C_F^{\theta_0}(\rho)$ will be infinite too, whereas physical models require a bounded derivative $\partial_\theta g_n^x(\theta)$ \cite{VSL2011,VSL2006}.

\end{document}